\documentclass[11pt]{article}

\usepackage{epsfig}
\usepackage{graphicx}
\usepackage[tight]{subfigure}
\usepackage{epstopdf}
\usepackage{amsfonts}
\usepackage{amssymb}
\usepackage{amsthm}
\usepackage{amsmath}
\usepackage{multirow}
\usepackage{array}

\newcommand{\newc}{\newcommand}
\newc{\ra}{\rightarrow}
\newc{\lra}{\leftrightarrow}
\newc{\be}{\begin{equation}}
\newc{\ee}{\end{equation}}
\newc{\bs}{\begin{split}}
	\newc{\es}{\end{split}}
\newc{\ba}{\begin{eqnarray}}
\newc{\ea}{\end{eqnarray}}
\newc{\ov}{\overline}
\newc{\pa}{\partial}
\newc{\D}{\Delta}

\newc{\nn}{\nonumber}
\newc{\bfe}{{\bf \ell}}
\newc{\alp}{\alpha}
\newc{\gam}{\gamma}
\newc{\bfeta}{{\bf \eta}}
\newc{\bfxi}{{\bf \xi}}
\newc{\del}{\delta}
\newc{\lam}{\lambda}
\newc{\sig}{\sigma}
\newc{\ups}{\upsilon}
\newc{\ome}{\omega}
\newc{\la}{\langle}
\newc{\rg}{\rangle}
\newc{\im}{\imath}
\newc{\rpv}{\slash{\hspace{-.2cm} R}}
\newc{\rpva}{\slash{\hspace{-.23cm} R}}
\begin{document}
	\begin{titlepage}

		\vspace*{0.7cm}

		\begin{center}
			{\Large {\bf $ n$-$\bar n$ oscillations and the neutron lifetime}}
			\\[12mm]
		George K. Leontaris$^{a}$
			\footnote{E-mail: \texttt{leonta@uoi.gr}} and
		John D. Vergados$^{a,b}$
			\footnote{E-mail: \texttt{vergados@uoi.gr}}
			\\[-2mm]
			
		\end{center}
		\vspace*{0.50cm}
		\centerline{$^{a}$ \it
			Physics Department, Theory Division, Ioannina University,}
		\centerline{\it
			GR-45110 Ioannina, Greece }
		\vspace*{0.50cm}
			\centerline{$^{b}$ \it Zhejiang University of 
		Technology (ZJUT), Hangzhou, Zhejiang, China}
		\begin{abstract}
			\noindent
			
	    Neutron-antineutron oscillations are considered in the light of  recently proposed  particle models, which claim to resolve the
	  neutron lifetime anomaly,  indicating the existence of baryon violating $\Delta B=1$ interactions. Possible constraints are derived
	  coming from  the non-observation of neutron-antineutron oscillations, which can take place if  the dark matter particle produced 
	  in neutron decay happens to be a Majorana fermion. It is shown that this can be realised 	in a simple MSSM extention where only
	  the baryon number violating term $u^cd^cd^c$ is included whilst all other R-parity violating terms are prevented to avoid rapid
	   proton decay.
	  It is demostrated how this scenario can be implemented in a string motivated GUT broken to MSSM by fluxes. 
		\end{abstract}
		
\end{titlepage}

\section{Motivation and facts}

Neutrons, together with protons and electrons are the fundamental constituents of atomic matter and their properties have been studied for almost a century.
A free neutron, in particular, disintegrates to a proton, an electron and its corresponding antineutrino, according to the well known $\beta$-decay 
process    $n\to p+e^-+\bar{\nu}_e$.  Notwithstanding those well known facts, the precise lifetime of the neutron remains a riddle  wrapped up in an enigma. 
The problem lies in the fact that the two distinct techniques employed to measure the lifetime end up to a  glaring discrepancy~\cite{Greene:2016SA}. More specifically, in one method a certain number of neutrons are   collected in a container \cite{Patrignani15} (known as ``bottle''), where, after a certain time duration   (comparable to the neutron lifetime), several of them decay. The remaining fraction of them can be used to determine the lifetime which is found to be  $\tau_n\approx 879.6\pm 0.6$ sec. In an alternative way 
of measuring the lifetime named \cite{ByrDaw96},\cite{YDGGLNSW13} ``beam'', a neutron beam with known intensity is directed  to an electromagnetic trap. Counting the emerging protons within a certain time interval, 
it is found that their numbers are consistent with  a neutron lifetime   of $\tau_n = 888.0\pm 2.0$ sec.  These two measurements display a $4.0\,\sigma$ discrepancy which cannot be attributed  to statistical uncertainties.   An explanation of this difference of the two measurements could be that other decay channels contribute to the total lifetime in the ``beam'' case.  In the context of the minimal Standard Model, however, there are no available   couplings and particles that lead to such a channel and could thus account for this difference.
According to a recent proposal~\cite{Fornal:2018eol} the discrepancy could be interpreted if neutrons have a decay channel to a dark matter (DM) candidate particle $\chi$
with a branching ratio $\sim 1\%$  and a mass comparable to the neutron's mass.  The simplest possibility is realised with the neutron decay to a two particle final state consisting of a DM fermion  $\chi$ and a monochromatic photon, $n\to \chi+\gamma$.  Operators describing this type of decays, however, violate baryon number. In the microscopic level, the  description of the above decay requires the existence of  a colour scalar field with the quantum numbers of a  Standard Model (SM) colour triplet, $D=(3,1)_{-1/3}$,
with mass $M_D\ge 1$ TeV  and  the couplings
\be
{\cal L}_{D\chi}\supset 
\lambda_q \overline{u^c}_L d_R D+ \lambda_{\chi}\bar{D}\bar{\chi}d_R+m_{\chi}\bar{\chi}\chi~\cdot
\label{Dchiterms}
\ee 
Two basic assumptions have been made for this scenario to work. Firstly, it is assumed that other baryon violating couplings of the new colour triplet, $D=(3,1)_{-1/3}$, 
are substantially suppressed. Indeed, a colour  triplet, introduces other baryon non-conserving couplings similar to the $R$-parity violating ($\rpva$) ones  of the supersymmetric theories. Unless their couplings are unnaturally tiny, they lead to fast proton decay at unacceptable rates. In the context of SM, there are no obvious symmetries which  prevent their appearance while leaving the terms~(\ref{Dchiterms}) intact. Secondly, 
it is assumed that the DM fermion   $\chi$ is a Dirac particle. 
Since  $\chi$ is a neutral field, however, it could be likewise a Majorana particle
and, in such a case, might contribute to $n$-$\bar n$ oscillations.

In this letter, we reconsider the  interpretation of the neutron lifetime discrepancy described above, in the context of Minimal Supersymmetric Standard Model (MSSM) extensions and in particular SUSY and String motivated GUTs. 
There are many good reasons to implement the above scenario in this context.  We firstly  remark that the kind of the scalar particle introduced to 
realise the processes has the quantum numbers of a down quark colour triplet. Thus, in the context of MSSM, this could be the scalar component 
$\tilde d^c_j$ of a down quark supermultiplet. There are good chances that the supersymmetry breaking scale is around  the  TeV scale  and the sparticle
spectrum may  be accessible either at the LHC or its upgrades. Thus, taking into account the recent bounds of LHC experiments,
its mass $m_{\tilde d^c_j}$ could be around  the TeV scale which is adequate to interpret the neutron lifetime discrepancy.
Notice however, that in the MSSM context, terms such as~(\ref{Dchiterms}) appear together with other baryon and lepton number violating interactions giving rise 
to fast proton decay, and therefore, they are forbidden by $R$-symmetry. There are examples of Grand Unified Theories with string origin, however, 
equipped with symmetries and novel symmetry breaking mechanisms where it is  possible to end up with a lagrangian  only with the desired  $\rpva$-coupling and all the others  forbidden.
Thus, in the presence only of the trilinear coupling shown in~(\ref{Dchiterms}) which can account for the discrepancy, the only possible baryon violating
processes is neutron-antineutron oscillations. Our aim in the present work, is to investigate under what conditions the issue of neutron lifetime is solved. 
In particular, we will examine whether 
the strength of the couplings and the mass scale required to interpret the discrepancy, are consistent with the bounds on $n$-$\bar n$ oscillations.

The layout of the paper is as follows:  In section 2 we present a short overview of gauge invariant baryon and lepton number violating symmetries in the context of GUTs with emphasis on $R$-parity violating supersymmetry. In section 3 we summarise the essential formalism related to neutron-antineutron oscillations and in section 4 we present the main results, including bounds of the relevant baryon violating couplings in the TeV scale  extracted from the current limits of $n$-$\bar{n}$ oscillations. Some concluding remarks and a short discussion are presented in section 5. Finally, for the readers convenience,   some detailed formulas  entering our  calculations  are given in the appendix.

\section{A brief overview of R-parity in fluxed GUTs}

In the non-supersymmetric standard model, at the renormalisable level, Baryon ($B$) and Lepton ($L$)  numbers are conserved quantum numbers,
due to accidental global symmetries.  This fact is consistent with the observed stability of the proton and the absence 
of lepton decays (such as $\beta\beta$-decay) which violate $B$ and $L$. Introducing new coloured particles which imply additional  interactions,
however, this is no longer true. 

In the supersymmetric lagrangian of the Standard Model symmetry, in principle, one could write down  gauge invariant
terms which violate $B$ and $L$  mumbers.  In superfield notation these are:
\be 
{\cal W}_{\rpv} \supset 
\lambda_{ijk} Q_i d^c_j \ell_k\,+\,\lambda'_{ijk} \ell_i \ell_j e^c_k\,+\, \lambda''_{ijk} u^c_i d^c_j d^c_k\,+\,\lambda_{h\ell}\,h_u\ell_j~\cdot \label{RPV} \; 
\ee 
If all these couplings were present, for natural values of Yukawas $\lambda_{ijk}\sim {\cal O}(10^{-1})$, violation of
$B$ and  $L$ would occur at unacceptable rates. 
As is well known, in the minimum supersymmetric standard model the adoption of $R$-symmetry prevents all these terms.

Without the existence of $R$-symmetry or other possible discrete and $U(1)$ factors, these terms are also present in GUTs. In the minimal $SU(5)$
for example, the most common $B$ and $ L$ violating terms arise from the coupling 
\ba 
10_f \cdot\bar 5_f\cdot\bar 5_f& \to &Qd^c\ell +e^c\ell\ell+u^cd^cd^c~\cdot \label{RPVsu5}
\ea
In a wide class of  string motivated GUTs  there are cases where some of the terms in~(\ref{RPVsu5}) are absent in 
a natural way.  In a particular class of such models, where the breaking of the gauge symmetry occurs due to fluxes which 
are switched on along the dimensions of the compact manifold, we may have  for example the following SM decomposition:
\ba 
10_f \cdot\bar 5_f\cdot\bar 5_f& \to &u^cd^cd^c +{\rm nothing\; else},
\ea
which is just the operator required to mediate $n$-$\bar n$ oscillations. The absence
of the remaining couplings in~(\ref{RPV}) ensures that the  proton remains stable, 
or its decay occurs at higher orders in perturbation theory and therefore its  decay rate is highly suppressed and undetectable from present day experiments. 

To be more precise,  focusing in $SU(5)$ as a prototype unified theory,
the flux mechanism works as follows~\cite{Beasley:2008dc}:
Assuming that $SU(5)$ chirality has been  obtained by fluxes associated with abelian factors embedded together with $SU(5)$ into 
a higher symmetry, another flux is introduced along the
hypercharge generator $U(1)_Y$ to break $SU(5)_{GUT}$~\cite{Beasley:2008dc}. It turns out that this is also responsible
for the splitting of the  $SU(5)$ representations.  If some integers   $M,N$ represent these two kinds of fluxes piercing 
certain  ``matter curves'' of the compact manifold hosting the 10-plets and 5-plets,
the following  splittings of the corresponding representations occur:
\ba
\#\underline{\bf 10}-\#\underline{\overline{\bf 10}}&\Rightarrow&
\left[\begin{array}{lcl}
	n_{{(3,2)}_{\frac 16}}-n_{{(\bar 3,2)}_{-\frac 16}}&=&M_{10}\\
	n_{{(\bar 3,1)}_{-\frac 23}}-n_{{(
			3,1)}_{\frac 23}}&=&M_{10}-N\\
	n_{(1,1)_{+1}}-n_{(1,1)_{-1}}& =&M_{10}+N\\
\end{array}\right.
\label{10j}
\\
\#\underline{\bf 5}-\#\underline{\overline{\bf 5}}&\Rightarrow&
\left[\begin{array}{lcl}
	n_{(3,1)_{-\frac 13}}-n_{(\bar{3},1)_{+\frac 13}}&=&M_{5}\\
	n_{(1,2)_{+\frac 12}}-n_{(1,2)_{-\frac 12}}& =&M_{5}+N\,\cdot\\
\end{array}\right.
\label{5i}
\ea
The integers $M_{10}, M_{5}, N$  are related to specific choices of the fluxes,
and  may take any positive or negative value,  leading to different number
of SM representations. 
Hence, there is a variety of possibilities which can be fixed only if certain string boundary conditions
have been chosen. In order to exemplify the effect of these choices, here we  assume only a few arbitrary cases where the integers $M, N$ 
take the lower possible values $\pm 1, 0$.\footnote{Of course, larger $M,N$ values are also possible. They may imply  different numbers
of SM representations on  matter curves	 but will not lead to new types of splittings~\cite{CrispimRomao:2016tww} other than those of Table~\ref{Table}.}
for the $SU(5)$ representations.  Substituting these numbers in~(\ref{10j},\ref{5i}) we obtain a variety of 
possibilities, and some of them are shown in Table~\ref{Table}.

\begin{table}[h!]
	\centering\begin{tabular}{cll|cll}
		\hline
		{ 10-plets}&Flux Units&\(10\) Content&{5-plets}&Flux Units &$\bar 5$ Content
		\\
		\hline
		$10_1$&$M_{10}=1, N={\hphantom +}0$ &$(Q,u^c,e^c)$ &$\bar 5_1$&\(M_{5}=+1, N={\hphantom +}0\)&$(d^c,\ell)$ \\
		$10_2$&$M_{10}=0, N=+1$ &$(-,\bar u^c,e^c)$ &$\bar 5_2$&\(M_5={\hphantom +}0, N=+1\)&$(-,\ell)$\\
		$10_3$&\(M_{10}=0, N=-1\) & $(-, u^c,\bar e^c)$  &$\bar 5_3$&\(M_5 ={\hphantom +} 0, N = -1\)&$(-,\bar \ell)$\\
		$10_4$&\(M_{10}=1, N=+1\) & $(Q,-,2 e^c)$ &$\bar 5_4$&\(M_5=+1, N=+1\)&$(d^c,2\ell)$\\
		$10_5$&\(M_{10}=1, N=-1\) & $(Q,2u^c,-)$ &$\bar 5_5$&\(M_5=+1, N=-1\)&$(d^c,-)$\\
		
		\hline
	\end{tabular}
	\caption{ Induced MSSM matter content  from fluxed $SU(5)$ representations~\label{Table}}
\end{table}
Hence, we  end up with incomplete $SU(5)$ representations.
Some examples of $R$-parity violating operators formed by trilinear terms involving the above  incomplete representations are shown in 
Table~\ref{Table2}.
(for a comprehensive analysis and a complete list of possibilities see~\cite{CrispimRomao:2016tww}).
\begin{table}[h!]\centering\begin{tabular}{llll}
		\hline
		\( SU(5)\)-term&MSSM content&$\slash{\hspace{-0.25cm}R}$-operator(s)&
		dominant process
		\\
		\hline
		$10_1 \bar 5_1\bar 5_1$&$(Q,u^c,e^c)(d^c,\ell)^2$
		&all&proton decay\\
		$10_1 \bar 5_4 \bar 5_4$&\((Q,u^c,e^c)(d^c,2\ell)^2\)
		&all&proton decay\\
		$10_1 \bar 5_3 \bar 5_3$& \((Q,u^c,e^c)(-,\bar \ell)^2\)
		&none&none\\
		$10_1 \bar 5_2 \bar 5_2$& \((Q,u^c,e^c)(-,\ell)^2\)
		&$\ell\ell e^c$&$\ell_{e,\mu,\tau}$ violation\\		
		$10_1\bar 5_5\bar 5_5$& \((Q,u^c,e^c)(d^c,-)^2\)&$u^cd^cd^c$&$n-\bar
		n$ oscillation\\
		$10_3\bar 5_5\bar 5_5$& \((-,u^c,\bar e^c)(d^c,-)^2\)&$u^cd^cd^c$&$n-\bar
		n$ oscillation\\		
		\hline
	\end{tabular}
	\caption{$SU(5)$-fluxed representations with incompleted MSSM content, 
		$\slash{\hspace{-.24cm} R}$-processes emerging from the trilinear coupling 
		$10_i\bar 5_j \bar 5_j$ for selected
		combinations of the  multiplets given in Table~\ref{Table}. \label{Table2}}
\end{table}
We observe that the couplings $10_1\cdot\bar 5_5\cdot \bar 5_5$ and  $10_3\cdot\bar 5_5\cdot \bar 5_5$ in the last two lines
of this table give exactly the required $R$-violating trilinear coupling, while all the
other couplings are absent. This is just the case that will be considered in the subsequent analysis.

\section{neutron-antineutron oscillation formalism}
In this section we will briefly present the main features of the $n$-$\bar{n}$ oscillations mainly to establish notation and put the recently baryon violating scenario, proposed for the extra exotic channel of neutron decay to a light dark matter, in a broader perspective. In this context  additional processes entering $n$-$\bar{n}$ oscillations at tree level or at the one loop level are presented.

\subsection{neutron and antineutron bound wavefunctions}
We will consider the neutron as a bound state  of three  quarks (antiquarks) for  neutron (antineutron), in a colour singlet $s$-state in momentum space. The orbital part is of the form:
\be
\Psi_{{\bf P}0s0s}({\bf Q},\bfxi,\bfeta)=\sqrt{3\sqrt{3}} (2 \pi)^{3/2}\delta(\sqrt{3}{\bf Q-P})\phi(\bfxi)\phi(\bfeta)~,
\ee
where ${\bf P}$ is the hadron momentum and:
\be
{\bf Q}=\frac{1}{\sqrt{3}}(\bf{p_1+p_2+p_3}),\,\bfeta =\frac{1}{\sqrt{6}}(\bf{p_1+p_2-2 p_3}),\,\bfxi =\frac{1}{\sqrt{2}}(\bf{p_1-p_2})~,
\ee
with $\bf{p }_i,\,i=1,2,3$ the quark momenta. 
The functions $\phi(\bfxi),\,\phi(\bfeta)$ are assumed to be   $0s$  harmonic oscillator wavefunctions. These functions are assumed to be normalised the usual way:
\be
\label{wfnorm}
\begin{split}
\langle \Psi_{P,0s,0s}|\Psi_{P'0s 0s}\rangle
&=(2 \pi)^{3}(3\sqrt{3})\int d^3{\bf Q}\delta(\sqrt{3}{\bf Q-P})\delta(\sqrt{3}{\bf Q-P'})\int d^3\bfxi|\phi(\bfxi)|^2 \int d^3\bfeta|\phi(\bfeta)|^2\nn\\
&=(2 \pi)^{3}\delta({\bf P-P'})~\cdot 
\end{split}
\ee

\subsection{neutron-antineutron transition mediated by dark matter Majorana fermion}

\begin{figure}[h!]
	\centering
	\subfigure[
	\label{Fig:nnbarMaja}]
	{\includegraphics[width=0.45\linewidth]{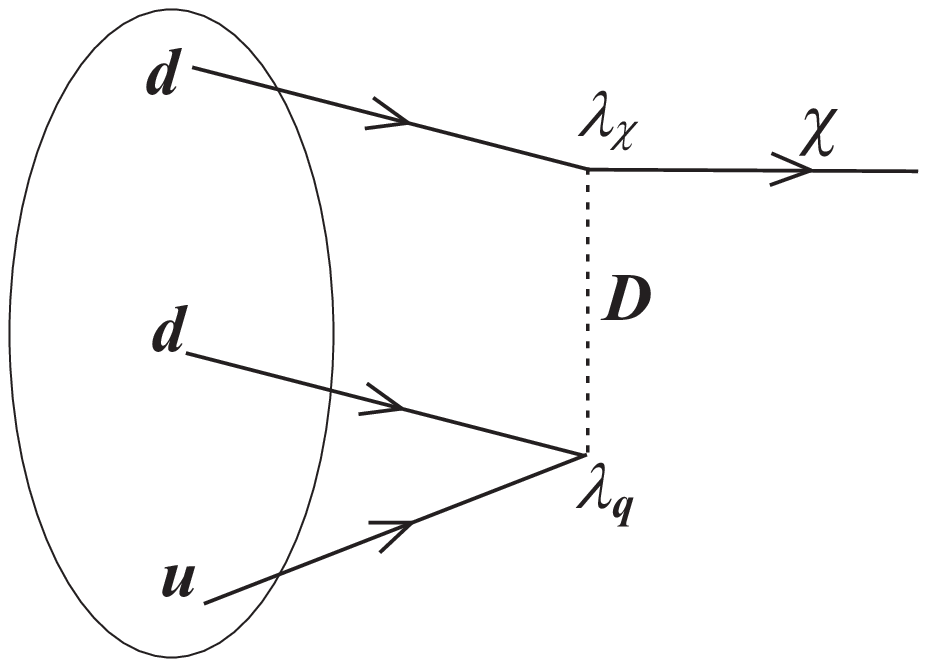}}
	\subfigure[
	\label{Fig:nnbarMajb}]
	{\includegraphics[width=0.45\linewidth]{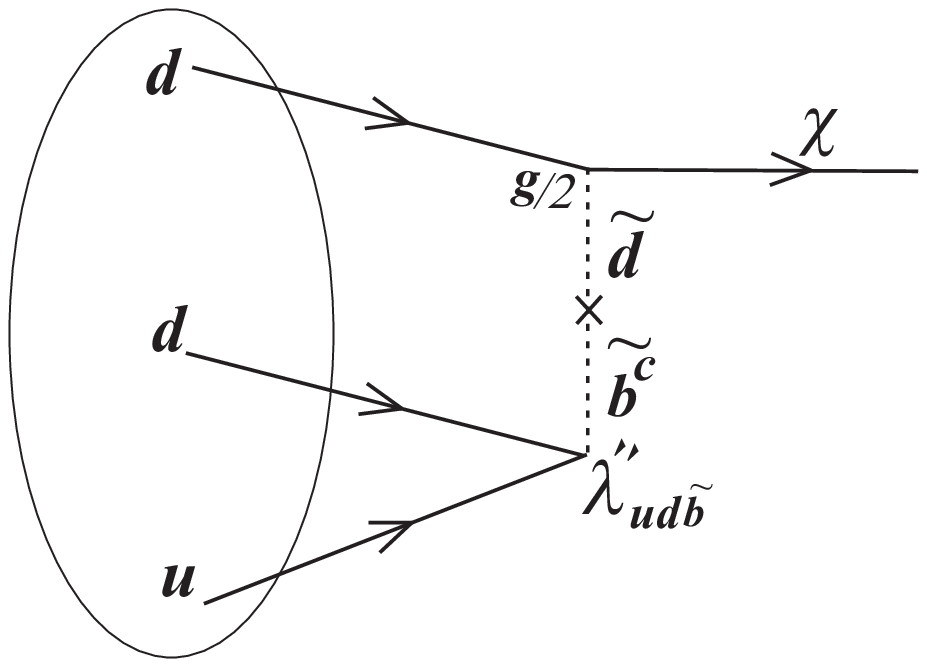}}
	\\
	\subfigure[
	\label{Fig:nnbarMajc}]
	{\includegraphics[width=0.45\linewidth]{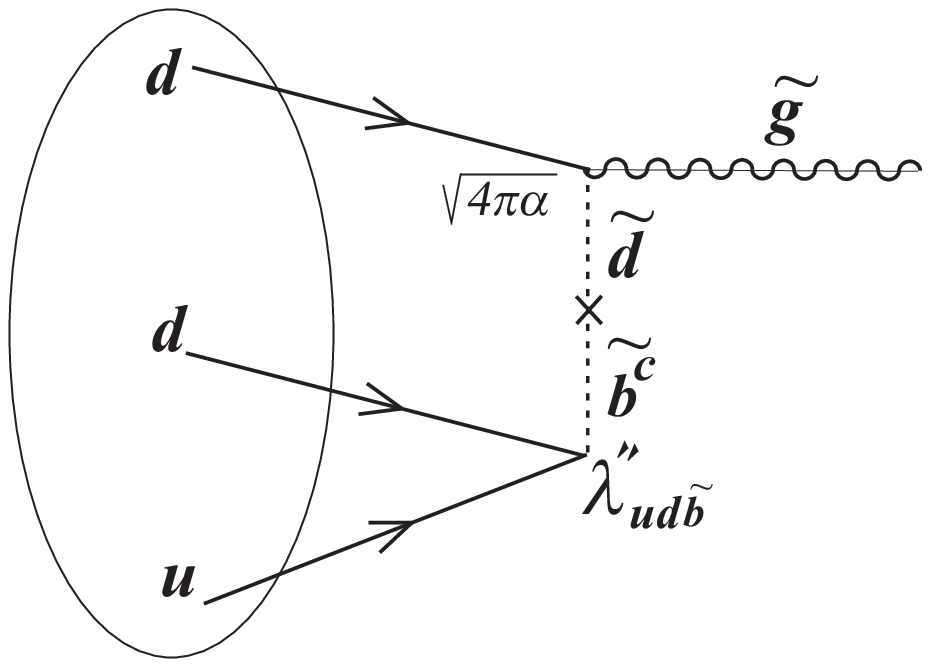}}
		\subfigure[
		\label{Fig:nnbarMajd}]
	{\includegraphics[width=0.45\linewidth]{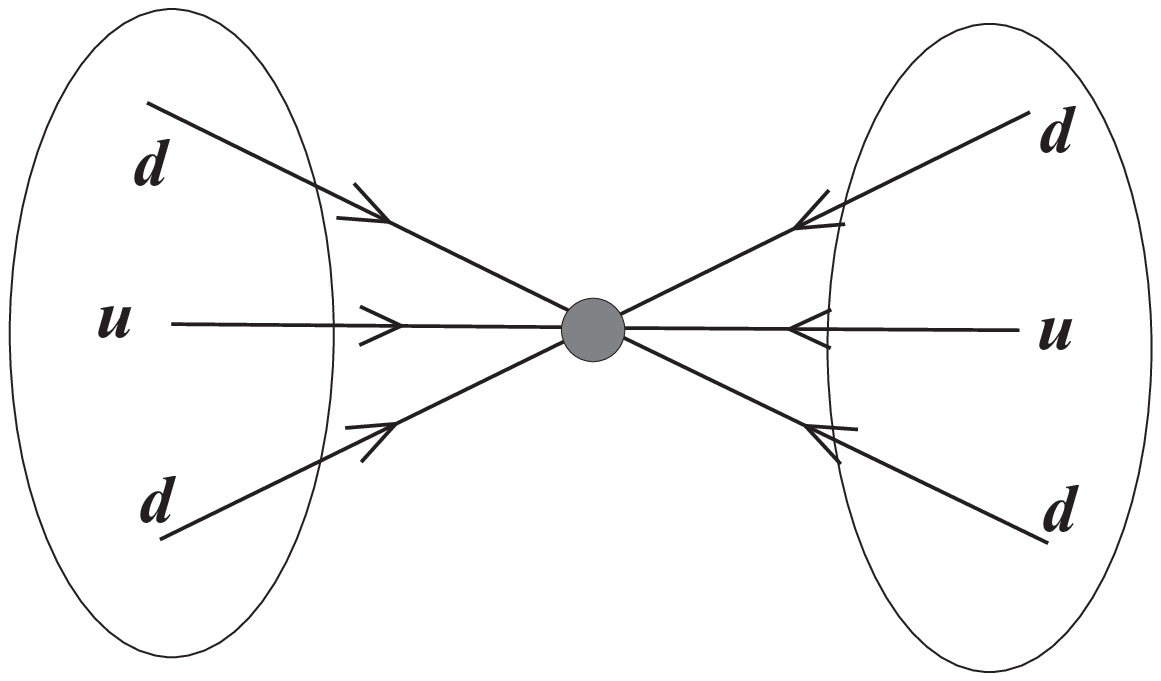}}
	\caption{(a)  A dark matter colourless Majorana fermion emitted from a neutron. (b) A dark matter particle $\chi$ emitted in the context of R-parity violating sypersymmetry, e.g. the  gaugino $\tilde{W}_3$ with coupling $g/2$ or $\tilde{B}$ with coupling $g'/2$ or any of the two Higgsinos with more complicated couplings. (c)  A gluino, emitted from a neutron in R-parity violating sypersymmetry.  The emitted Majorana fermion  propagates  and it can get absorbed by  an antineutron leading to $n$-$\bar{n}$  oscillations. (d) Such an oscillation can also be  induced by a box  diagram leading to a contact interaction.}
\end{figure}

A dark matter colourless Majorana fermion of mass $m_{\chi}$  emitted from a neutron of momentum $P$  and absorbed by  an antineutron of momentum $P'$ can lead to $n$-$\bar n$ oscillations. This process   is exhibited in Fig.~\ref{Fig:nnbarMaja}. In a previous study~\cite{Fornal:2018eol} this did not happen, since the mediating fermion was assumed to be a Dirac like particle, but there is no reason to restrict in this choice. In fact there exists the possibility of this particle being  a Majorana like fermion in which case neutron-antineutron  oscillations become possible.

The orbital matrix element takes the form:
\ba
ME&=&(2 \pi)^{3}(3\sqrt{3})\left (\frac{g_q g_{\chi}}{m^2_{D}}\right )^2\frac{1}{m_{\chi}}\int d^3{\bf Q}\int d^3{\bf Q'}\delta(\sqrt{3}{\bf Q-P})\delta(\sqrt{3}{\bf Q'-P'})\delta(\sqrt{3}({\bf Q-Q'})\nn\\ 
& &\int d^3\bfxi \int d^3\bfeta\int d^3\bfxi'\int d^3\bfeta'\phi(\bfxi)\phi(\bfeta)\phi(\bfxi') \phi(\bfeta')~\cdot 
\ea
Using (\ref{wfnorm}) the matrix element can be written as follows:
\ba
ME&=&(2 \pi)^{3}\delta{(\bf{P-P'})}{\cal M}_{orbital},\nn\\ {\cal M}_{orbital}&=&\frac{1}{3\sqrt{3}}\left (\frac{g_q g_{\chi}}{m^2_{D}}\right )^2\frac{1}{m_{\chi}}
\int d^3\bfxi \int d^3\bfeta\int d^3\bfxi'\int d^3\bfeta'\phi(\bfxi)\phi(\bfeta)\phi(\bfxi') \phi(\bfeta')~\cdot 
\ea
Now the $0s$ wavefunction is
$$
\phi({\bf x})=\left (\frac{b_N}{\sqrt{\pi}}\right )^{3/2}e^{-\frac{b^2_N x^2}{2}},\,{\bf x}=\bfxi,\bfeta,\bfxi',\bfeta'. $$
Thus, performing the Gaussian integral, we get
$$
I=\int d^3{\bf x}\phi({\bf x})=\left (\frac{b_N }{\sqrt{\pi}}\right )^{3/2} 4 \pi\int_0^{\infty}dx\,  x^2e^{-\frac{b^2_N  x^2}{2}} =2 \sqrt{2} \left (\frac{\sqrt{\pi}}{b_N }\right )^{3/2}~,$$
and the orbital part  becomes
\be
{\cal M}_{\mbox{\tiny{orb}}}=\frac{64}{3\sqrt{3}}\pi^3\left (\frac{\lambda_q \lambda_{\chi}}{m^2_{D}}\right )^2\frac{1}{m_{\chi}}\left (\frac{1}{b_N }\right )^{6}~\cdot 
\ee
It is instructive to compare this with the probability for finding the quark at the origin inside the nucleon:
$$ |\psi(0)|^2=\frac{1}{\pi\sqrt{\pi}}\frac{1}{b_N^3}. $$
Then
\be
{\cal M}_{\mbox{\tiny{orb}}}=\frac{64}{3\sqrt{3}}\pi^6\left (\frac{\lambda_q \lambda_{\chi}}{m^2_{D}}\right )^2\frac{1}{m_{\chi}}|\psi(0)|^4~\cdot 
\label{Eq:exprphi}
\ee

The colour factor is quite simple since it involves  the same hadron. It takes the form:
\be
\sum_{\alpha}(ud)_S(0,1)_{-\alpha}(-1)^{\phi_{\alpha}}d(1,0){\alpha} =\sqrt{3}(0,0)~,
\ee
where $(ud)_S(0,1)_{-\alpha} $ is the flavor symmetric colour antisymmetric two quark state, $\phi_{\alpha}$ the  conjugation phase, and (0,0) is the colour singlet hadronic state. Thus 
\be
{\cal M}_{\mbox{\tiny{colour}}}=3~,
\ee
\be
{\cal M}_{DM}=\frac{64}{\sqrt{3}}\pi^3\frac{1}{2}\left (\frac{\lambda_q \lambda_{\chi}}{m^2_{D}}\right )^2\frac{1}{m_{\chi}}\left (\frac{1}{b_N }\right )^{6}~\cdot
\ee
The factor of $\frac{1}{2}$ came from chirality since the propagating fermion is only left-handed.

In the case of supersymmetry induced oscillation, Fig.~\ref{Fig:nnbarMajb}, we find an analogous expression. The orbital part is similar to the previous one with the obvious modifications $m_{D}\rightarrow m_{\tilde{d}}$, $\lambda_q \lambda_{\chi}\rightarrow c_{bd}\lambda''_{ub\tilde d}\, g/2$. Here, $c_{bd}$ is the flavor violating  mixing between the scalars $\tilde{d}$ and $\tilde{b^c}$, which induces baryon violation. Thus:
\be
{\cal M}_{\mbox{\tiny{orbital}}}=\frac{64}{3\sqrt{3}}\pi^3\left ( c_{b\tilde{d}}\frac{\lambda''_{ud\tilde{b}}\, g/2}{m^2_{\tilde{d}}} \right )^2 \frac{1}{m_{\tilde{W}_3}} \left (\frac{1}{b_N }\right )^{6}~\cdot 
\label{Eq:SUSYDM}
\ee
Including the colour helicity factors we set
\be
{\cal M}_{{\tiny SUSYDM}}=\frac{64}{\sqrt{3}}\pi^3\frac{1}{2}\left (c_{b\tilde{d}}\frac{\lambda''_{ud\tilde{b}}\, g/2}{m^2_{\tilde{d}}} \right )^2 \frac{1}{m_{\tilde{W}_3}} \left (\frac{1}{b_N }\right )^{6}~\cdot 
\ee

\subsection{Additional neutron-antineutron mechanisms at tree level }

\begin{figure}[htb!]
	\begin{center}
			\includegraphics[width=0.6\textwidth,height=0.3\textwidth]{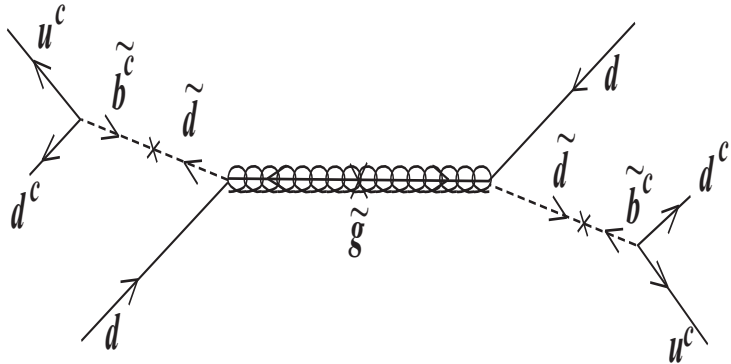}
	\end{center}
	\caption{\it $n$-$\bar n$ oscillations with Gluino exchange takes place at tree level. } \label{gluino}
\end{figure}

 $n$-$\bar n$ oscillations with gluino exchange take place at tree level, see fig~\ref{gluino}. 
This is directly comparable with $n$-decay process through DM particle $\chi$.
 However, because of the colour antisymmetry, the coupling $u^cd^c\tilde d^c$ cannot be realised directly and it requires 
mass insertion, thus a suppression factor emerges due to assumed mixing  between $\tilde b^c, \tilde s^c$ and their
left components $\tilde b, \tilde s$. (This requirement is beyond the minimal flavour violation scenario which assumes
diagonal mass matrix.)

It is known that  {\it di-nucleon}  decays to two Kaons, $N\,N\to K\,K$, impose stringent constraints on the  coupling $\lambda''_{ud\tilde s}$~\footnote{See for example~\cite{Goity:1994dq,Calibbi:2016ukt} and references therein.}. 
Hence, we will focus only on $\lambda''_{ud\tilde b}$ which becomes more relevant for neutron-antineutron oscillations. 

\noindent
The gluino exchange diagram of Fig.~\ref{Fig:nnbarMajc}, (see also Fig.~\ref{gluino}), differs from that of Fig.~\ref{Fig:nnbarMajb} in the sense that  the gluino is a colour octet and interacts strongly. Thus
\be
{\cal M}_{\mbox{\tiny{orbital}}}=\frac{64}{3\sqrt{3}}\pi^3\left ( c_{b\tilde{d}}\frac{\lambda''_{ud\tilde{b}} \sqrt{4 \pi \alpha_s}}{m^2_{\tilde{d}}} \right )^2\frac{1}{m_{\tilde{g}}}\left (\frac{1}{b_N }\right )^{6}~\cdot 
\ee
The colour factor is a bit more complicated. We encounter the combination:
\be
\sum_{\alpha,\beta,\gamma}(ud)_S(01)_{\alpha}d_{\beta}\tilde{g}_{\gamma} c_{\alpha,\beta,\gamma}~,
\ee
with a similar combination on the other hadron. The states are specified as follows:
$$\alpha=1\Leftrightarrow |(0,1)-2,0,0\rangle,\,\alpha=2\Leftrightarrow |(0,1)1,\frac{1}{2},-\frac{1}{2}\rangle,\,\alpha=3\Leftrightarrow |(0,1)1,\frac{1}{2},\frac{1}{2}\rangle$$
$$\beta=1\Leftrightarrow |(1,0)2,0,0\rangle,\,\beta=2\Leftrightarrow |(1,0)-1,\frac{1}{2},\frac{1}{2}\rangle,\,\beta=3\Leftrightarrow |(1,0)1,\frac{1}{2},-\frac{1}{2}\rangle$$
$$\gamma=1\Leftrightarrow |(1,1)3,\frac{1}{2},\frac{1}{2}\rangle,\,\gamma=2\Leftrightarrow |(1,1)3,\frac{1}{2},-\frac{1}{2}\rangle,\,\gamma=3\Leftrightarrow |(1,1)0,1,1\rangle,\,\gamma=4\Leftrightarrow |(1,1)0,1,0\rangle$$ $$\gamma=5\Leftrightarrow |(1,1)0,1,-1\rangle,\,\gamma=6\Leftrightarrow |(1,1)0,0,0\rangle,\,\gamma=7\Leftrightarrow |(1,1)-3,\frac{1}{2},\frac{1}{2}\rangle,\,\gamma=8\Leftrightarrow |(1,1)-3,\frac{1}{2},\frac{1}{2}\rangle,$$
in the standard SU(3) labeling of the states~\cite{Hecht:1965} $|(\lambda,\mu)\epsilon,\Lambda,\Lambda_0\rangle$.

The allowed by SU(3) symmetry coefficients can be easily calculated from the tables involving the reduction $(01)\otimes (10)\rightarrow (11) $, table 1 of ref.~\cite{Vergados:1968sha}. The obtained  results are presented in table~\ref{tab:SU(3)colour}.
\begin{table}[!t]
	\begin{center}
		\caption{The non vanishing coefficients $c_{\alpha,\beta,\gamma}$ allowed by the SU(3) symmetry. For notation see text.
			\label{tab:SU(3)colour} }
		\begin{tabular}{|ccc|c|}
			\hline
			\hline
			$\alpha$&$\beta$&$\gamma$&$c_{\alpha,\beta,\gamma}$\\
			\hline
			3&1&1&1\\
			2&1&2&1\\
			3&2&3&1\\
			2&1&2&1\\
			3&3&4&$\frac{1}{\sqrt{2}}$\\
			2&2&5&$-\frac{1}{\sqrt{2}}$\\
			1&1&6&$\frac{\sqrt{2}}{\sqrt{3}}$\\
			3&3&6&$-\frac{1}{\sqrt{6}}$\\
			2&2&6&$\frac{1}{\sqrt{6}}$\\
			1&2&8&1\\
			1&3&8&1\\
			\hline
			\hline
		\end{tabular}
	\end{center}
\end{table}
Expanding the hadronic states in terms of an antisymmetric pair of quarks and a single quark we find
\be
{\cal M}_{\mbox{\tiny{colour}}}=-\frac{1}{\sqrt{3}}\sum_{\alpha,\beta,\gamma}3 (c_{\alpha,\beta,\gamma})^2\left (-\frac{1}{\sqrt{3}}\right )=8~,
\ee
and thus, we get:
\be
{\cal M}_{\mbox{\tiny{gluino}}}=\frac{512}{3\sqrt{3}}\pi^3\frac{1}{2}\left (
c_{b\tilde{d}}\frac{\lambda''_{ud\tilde{b}} \sqrt{4 \pi \alpha_s}}{m^2_{\tilde{d}}} \right )^{6}~\cdot 
\ee

\subsection{neutron-antineutron transition mediated by box diagrams}
\begin{figure}[htb!]
	\begin{center}
	\includegraphics[width=0.6\textwidth,height=0.3\textwidth]{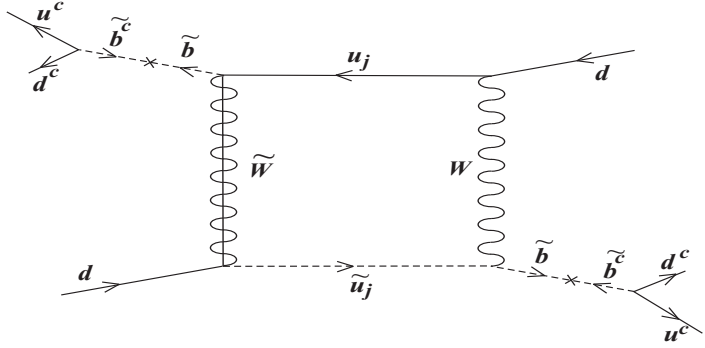}
	\end{center}
	\caption{\it $n-\bar n$ oscillations from box contributions. } \label{box}
\end{figure}
In this case there is no need to have flavour off diagonal baryon violating interactions, a mixing between the scalars $\tilde{b}$ and $\tilde{b^c}$ is adequate. The generation mixing can be induced as in the standard model via the Wino and the W-boson in a box diagram. In this case the interaction between the neutron and antineutron does not take the simple form found above at tree level. Since, however, the $\tilde b$-scalars are quite heavy, it leads to a contact interaction, see Fig. \ref{Fig:nnbarMajd}. Since no colour particle propagates between the two hadrons the colour factor is 3 and the orbital part can be written in the form:
\be
{\cal M}_{\mbox{\tiny{orbital}}}=\frac{64}{3\sqrt{3}}\pi^3\left ( c_{b\tilde{b}}\lambda''_{ud\tilde{b}} \right )^2\frac{1}{m^4_{\tilde{d}}} s_{{\tiny {box}}} \frac{1}{m_{\tilde{W}}}\left (\frac{1}{b_N }\right )^{6}~,
\ee
where $s_{{\tiny {box}}}$ is dependent on the masses of the particles circulating in the loop, namely the $W$-boson, the top quark, the Wino and the $\tilde{u}$ scalars. Thus
\be
{\cal M}_{\mbox{\tiny{box}}}=\frac{64}{3\sqrt{3}}\pi^3\frac{3}{2}\left ( c_{b\tilde{b}}\lambda''_{ud\tilde{b}} \right )^2\frac{1}{m^4_{\tilde{d}}}g^4 s_{{\tiny {box}}}\frac{1}{m_{\tilde{W}}}\left (\frac{1}{b_N }\right )^{6},
\label{Eq:BoxExpr}
\ee
where $g^2=4\sqrt{2}G_Fm^2_W\approx0.4$ and $s_{{\tiny {box}}}$ will be evaluated in Appendix A.
\section{neutron-antineutron oscillation results}
\label{sec:Results}

Combining the two cases, namely the non-supersymmetric dark matter and  the corresponding 
supersymmetric  processes, we find that   the transition amplitude takes the form:
\\

\ba
{\cal M}&=&m_{n}\kappa,\,\kappa=\frac{64}{3\sqrt{3}}\pi^3\frac{1}{2}\left[3\left ( \left (\frac{\lambda_q \lambda_{\chi}}{b_N^22m^2_{D}}\right )^2\frac{1}{b_N^2 m_n m_{\chi}}\right )+3 \left(  \frac{ c_{b\tilde{d}}\lambda'' _{ud\tilde{b}}\, g/2}{b^2_N m^2_{\tilde{d}}}\right )^2\frac{1}{b_N^2m_n m_{\tilde{W}_3}}\right .\nonumber\\&+&\left . 8 \left(  \frac{ c_{b\tilde{d}}\lambda'' _{ud\tilde{b}} \sqrt{4 \pi \alpha_s}}{b^2_N m^2_{\tilde{d}}}\right )^2\frac{1}{b_N^2m_n m_{\tilde{g}}}+
3\left (\frac{ c_{b\tilde{b}}\lambda''_{ud\tilde{b}} }{b_N^2m^2_{\tilde{d}}}\right )^2\frac{g^4 s_{{\tiny {box}}}}{b_N^2 m_n m_{\tilde{W}}}
+ \cdots \right  ]
\ea
where $s_{{\tiny {box}}}\approx 3.0 \times 10^{-6} $, see  the Appendix A. Due to this factor as well as the small mixing  $c_{b\tilde{b}}$, the parameter $\lambda''_{ud\tilde{b}}$ need not be extremely small. Notice also that a graph involving the bino will give a contribution similar to the second term.
Analogous graphs involving Higgsinos are also possible, but they provide  no new insights and will not be elaborated.

It is now natural to assume that the mass of the propagating scalar is the same in all models. If constraints come from other experiments we will compensate by adjusting the relevant couplings. Then we can take as scale of the masses to be of the order 1 TeV. Another parameter to be determined is the nucleon size parameter which is usually taken to be 0.8 fm. This is related to the nucleon wavefunction at the origin 
$$ \psi(0)^2=\frac{1}{\pi \sqrt{\pi}}\frac{1}{b_N^3}~\cdot $$
In reference \cite{Fornal:2018eol} the value of $\psi(0)^2=0.014$ GeV$^3$ was adopted taking into account effects arising from lattice gauge calculations \cite{LatticeGauge17}. This leads to a value of about 0.5 fm. We will adopt this value in the present calculation.  
Thus we can  write $\kappa$ in the form:
\be
\kappa= \kappa_0 \kappa_1,\; \;\kappa_0=4.0 \times 10^{-15} ,\label{kappas}
\ee 
with 
\be 
\kappa_1=\left[3 \left (\lambda_q \lambda_{\chi}\right )^2\frac{m_n} m_{\chi}+3 \left(  c_{b\tilde{d}}\lambda'' _{ud\tilde{b}} \,g/2\right )^2\frac{m_n} {m_{\tilde{W}_3}} +8 \left(  c_{b\tilde{d}}\lambda'' _{ud\tilde{b}} \sqrt{4 \pi \alpha_s}\right )^2\frac{m_n} {m_{\tilde{g}}} +3\left ( c_{b\tilde{b}}\lambda''_{ud\tilde{b}} \right )^2 g^4 s_{{\tiny {box}}}\frac{ m_n}{ m_{\tilde{W}}} \right ]~.\label{k1}
\ee

The $n$-$\bar{n}$ mixing matrix becomes
$$
m_n\left( \begin{array}{cc}1&\kappa\\\kappa& 1 \\ \end{array}\right )~,
$$
which leads to complete mixing with energies $E_1=m_n(1+\kappa)$, $E _2= m_n(1-\kappa)$. Thus the neutron-antineutron oscillation probability in vacuum becomes
\be
P(n\leftrightarrow \bar{n})=\frac{1}{2}\left |e^{i E_1 t}-e^{-i E_2 t}\right |^2=\sin^2{(m_n\kappa t)}~\cdot 
\ee
In other words the oscillation time is:
\be
\tau=\frac{1}{m_n\kappa} \approx \frac{7 \times 10^{-24}}{\kappa} \mbox{s}\approx \frac{1.8 \times 10^{-9}}{\kappa_1} \mbox{s}~\cdot 
\ee
In the presence of matter the diagonal elements of the matrix are not the same, since the neutron and the antineutron interact differently with any surrounding  magnetic field or matter. A tiny magnetic field of the order of $10^{-10}$ T can lead to an energy difference of $ \sim 10^{-26}m_n$. In  current experiments the  magnetic fields are limited \cite{Phillips:2014fgb} to $10^{-8}$ T, which leads  $\delta\approx 10^{-24}$. Thus the oscillation probability becomes \cite{Phillips:2014fgb}
\be
P(n\leftrightarrow \bar{n})=\frac{(\kappa/\delta)^2}{1+(\kappa/\delta)^2}\sin^2{(m_n\sqrt{\delta^2+\kappa^2}t)}e^{-\lambda t}\approx \frac{\kappa^2}{\delta^2} \sin^2{(m_n\sqrt{\delta^2+\kappa^2}t)}\,e^{-\lambda t}~\cdot 
\ee
where $\lambda=1/\tau_n$ with $\tau_n$  the neutron life time, as e.g., measured in the ``bottle'' experiment mentioned in the introduction.
The non-observation of neutrino oscillations implies $\kappa\ll \delta$, $\kappa<10^{-9}\delta= 10^{-33}$, $\tau>10^8$ s.

In the dark matter mediated process  the value of  $\kappa_1=3 \left (6.7 \times 10^{-6} \right )^2\approx 1.3 \times 10^{-10}$  was employed \cite{Fornal:2018eol}. Thus
\be
\tau\approx 1.3\times 10^{1}\mbox{s},\,\kappa\approx 5\times  10^{-25}~\cdot 
\ee
This is in conflict with the neutron oscillation data.

In the context of the $R$-parity violating supersymmetry model~\cite{Goity:1994dq}  we will try to extract a limit on the value of $c_{b\bar{d}}\lambda _{ud\tilde{b}}$ in the case of the tree diagrams and the $c_{b\bar{b}}\lambda _{ud\tilde{b}}$ in the case of the box diagram
from the non observation of $n$-$\bar{n}$ oscillation, i.e., solve the relations:
\ba
&&\left[3\left(  c_{b\tilde{d}}\lambda _{ud\tilde{b}}\,g/2 \right )^2\frac{m_n} {m_{\tilde{W}_3}} +8 \left(  c_{b\tilde{d}}\lambda _{ud\tilde{b}} \sqrt{4 \pi \alpha_s}\right )^2\frac{m_n} {m_{\tilde{g}}} \right . \nonumber\\&& \left . 3 \left(  c_{b\tilde{d}}\lambda _{ud\tilde{b}} \sqrt{4 \pi \alpha}\right )^2\frac{m_n} {m_{\tilde{W}}}+3\left ( c_{b\tilde{b}}\lambda_{ud\tilde{b}} \right )^2 g^4 s_{{\tiny {box}}}\frac{ m_n}{ m_{\tilde{W}}} \right ]\le 10^{-33}/\kappa_0= 2.5 \times 10^{-17}
\ea
We will consider each case separately:
\begin{itemize}
\item Gluino exchange. Taking $\alpha_s=1$ and $m_{\tilde{g}}=500\,$GeV we obtain:
\be
|c_{b\tilde{d}}\lambda _{ud\tilde{b}}|\;\lesssim \; 1.2 \times 10^{-8}~\cdot 
\ee
\item A SUSY dark matter particle ($\tilde{W}_3$ or $\tilde B$), exchange (see fig. \ref{Fig:nnbarMajd}(b)). Taking  $m_{\tilde{W}_3}=500\,$GeV we obtain:
\be
|c_{b\tilde{d}}\lambda _{ud\tilde{b}}|\;\lesssim\; 2.0 \times 10^{-7}~\cdot 
\ee
\item Finally in the case of the box diagram taking $s_{{\tiny {box}}}= 3.0\times 10^{-6}$  and $m_{\tilde{W}}=500$ GeV we obtain a weaker upper bound of the order
\be
|c_{b\tilde{b}}\lambda _{ud\tilde{b}}|\;\lesssim \;     10^{-4}~\cdot 
\ee
\end{itemize}

\section{Discussion}

In a recent paper~\cite{Fornal:2018eol} a very interesting proposal was made to resolve the long standing discrepancy on the determination of neutron lifetime 
measured in experiments involving  trapped   neutrons in a ``bottle''  and neutrons decaying in flight (``beam'' experiments). This model considers  novel mechanisms
 for neutron decays involving new dark decay channels  in  the ``bottle'' case, where the decay products contain light dark matter particles, with mass in a slim range 
 between the neutron and proton mass. The final state of this reaction might also involve  visible particles such as photons. 
  These scenarios sparked off a renewed  activity   on this issue and astrophysical as well as experimental constraints on the various decay modes  have been discussed.  Hence,   in a recent analysis~\cite{Tang:2018eln} decay channels involving a light dark matter  particle and a visible photon were ruled out,   while decays involving dark photons are subject to stringent constraints from astrophysical observations~\cite{Cline:2018ami}.  Furthermore, 
  it has been suggested~\cite{Motta:2018rxp} (see also~\cite{Baym:2018ljz})
 that neutron decay to dark matter is in conflict with neutron stars, but the argument  does not involve free neutrons.
 
 In the present work, we have explored  two different aspects of this proposal, namely the implications on baryon number   violation and  the possible Majorana nature of the emitted dark matter light particle.

We firstly focused on the fact that this   decay process is realised with the mediation of  colour triplets. In the context of the Standard Model and its obvious supersymmetric extensions, such particles generate other dangerous baryon and lepton number violating interactions, unless their coupling strengths to ordinary 
matter are unnaturally small. We have suggested that this problem can be remedied in the case of a class of  SUSY GUTs derived in the 
framework of string theories where ``fluxes'' developed along the compact dimensions are capable of eliminating the  superpotential terms 
associated with the undesired interactions. 

Furthermore, we have considered the possibility that the neutral dark matter particle in the putative exotic neutron decay
channel is of  Majorana type. 
In this case we find, however, that the parameters employed in this model are in conflict with the neutron-antineutron oscillation limits. We have considered 
limits from such baryon number violating processes in the context of $R$-parity violating supersymmetry, both  at tree as well as at the one-loop order. We find the most stringent limit on the parameter $|c_{b\tilde{d}}\lambda _{ud\tilde{b}}|\le 1.2 \times 10^{-8}$ comes from gluino exchange. The weaker limit of $ |c_{b\tilde{b}}\lambda _{ud\tilde{b}}|\lesssim   10^{-4}$ comes from the box diagram. The difference can be attributed to the fact that the tree diagrams involve both baryon and family flavor change of the participating s-quarks, while the loop diagram is diagonal in flavor.

\vspace{2cm}
   GKL would like to thank Maria Vittoria Garzelli at UDEL for useful  suggestions.

\newpage
\section{Appendix A: The box contribution}

For the non expert reader we provide some details  regarding the evaluation of the box diagram contribution.

\noindent
The gluino exchange diagram requires non-minimal mixing which might not be present in simple supersymmetric models.  
Hence,  we assume the case where $\tilde b_L, \tilde b_R$ have a non-trivial mixing term 
\be 
m^2_{b_{L,R}}= m_b A_{eff}~, \label{mLR}
\ee 
where $A_{eff}=A-\mu \tan\beta $, $A$ being a soft SUSY breaking parameter, $\mu$  the Higgs mixing ($\mu$-term)
and $\tan\beta $ the Higgs vev ratio.  Similar terms can exist for the other two families. 
As a result,  the process receives contributions   from  one-loop box graphs involving Winos.  This  is depicted in fig~\ref{box}.  The possible $R$-parity violating terms  contributing  to the process are  $\lambda''_{ud\tilde b} u^c d^c b^c$ and $\lambda''_{ud\tilde s} u^c d^c s^c$. Only $\lambda''_{ud\tilde b}$ is shown in the figure since, as
explained above, $\lambda''_{ud\tilde s} $ is suppressed. Moreover, due to the larger $b$-quark mass $m_b$  compared to $m_s$, factors such as $m_{b}^{2}/m_{W}^2$,   enhance the effect. 

\noindent
The processes requires the sequence of  reactions: Initially   $d_R\,u_R+ d_L\,\ra\,d_L+\tilde{b}_{R}^*$
followed by  $d_L+\tilde{b}_{L}^*\ra \tilde{b}_L+{\bar{d}_L}$, from the $W$-boson and Wino exchange box diagram.  At the final stage we get   $\bar{d}_L+\bar{d}_R\,\bar{u}_R$. 

\noindent
Calculation of the diagram gives the following relation for the decay rate~\cite{Goity:1994dq}
\ba
\label{GammaW}
\Gamma=-\frac{({\lambda''_{ud\tilde b}})^2 g^4m_{\tilde{b}_{LR}}^{4}
	m_{\tilde{w}}}{32 \pi^{2}(m_{\tilde{b}_{L}}m_{\tilde{b}_{R}})^{4}}|\psi(0)|^{4}\sum_{j,k=1}^{3}\xi_{jk}\,\Omega (m_{\tilde{w}}^{2},m_{W}^{2},m_{u_{j}}^{2},m_{\tilde{u}_{k}}^{2}) ,
\ea
with $m^2_{\tilde{b}_{LR}}$ given by~(\ref{mLR}) and $\xi_{jk}$ being the following combination of  CKM matrix parameters: 
\be  
\xi_{jk}\,=\,V_{bj}\, V_{jd}^{\dagger}\, V_{bk}\, V_{kd}^{\dagger}~\cdot\label{CKMxng}
\ee 
The computation of the loop integral in~(\ref{GammaW}) is parametrised by the  function $\Omega$ which depends on the four masses circulating in the box and is  given by:
\be
\Omega (m_1,m_2,m_3,m_4)=\sum_{j=1}^4
\frac{m_j^{4}\ln(m_j^2)}{\prod_{k\ne j}(m_j^2-m_k^2)}~\cdot
\ee
The  current experimental lower bound on $n$-$\bar{n}$ oscillation period  
$\tau=\frac{1}{\Gamma}$ is  $\tau\,\gtrsim\, 10^8$s~\cite{Phillips:2014fgb}, (see section \ref{sec:Results}).

In our notation   	 $|\psi(0)|^2$ is the baryonic wavefunction matrix element  for three quarks inside a nucleon
estimated~\cite{Fornal:2018eol,LatticeGauge17} to be  $|\psi(0)|^2=0.014$ GeV$^3$.
From eq.( \ref{GammaW}) we can  recalculate the bounds on $\lambda''_{udb\tilde b}$ 
coupling using the latest LHC bounds on scalar masses involved in the box graph. 
However, knowing only the lower bounds on this large number of arbitrary mass
parameters through this complicated formula, is not very  illuminating. 
Thus, before going to the most general case, in order to reduce the number of arbitrary mass parameters, and  have a feeling of the 
contributions of the various components, we first examine the limit $m_{\tilde u}\to m_{\tilde W}$ 
and $m_{u,c}\ll m_{\tilde u}$. Then the various contributions of the integral become
simpler. In particular, 
those involving only the CKM mixing of the first two generations are simplified as 
follows 
\ba  
\Omega_{ij}&=&\frac{m_{\tilde{u}}^2-m_W^2-m_W^2 \log
	\left(\frac{m_{\tilde{u}}^2}{m_W^2}\right)}{\left(m_{\tilde{u}}^2-m_W^2\right){}^2}
\approx {\cal O}\left(\small{\frac{1}{2m_W^2}}\right),\; i,j=1,2~\cdot\label{12gen}
\ea 
The remaining contributions  become
\ba 
\Omega_{i3}&=&\frac{m_{\tilde{t}}^2 \log \left(m_{\tilde{t}}^2\right)}{\left(m_{\tilde{t}}^2-m_{\tilde{u}}^2\right)
	\left(m_{\tilde{t}}^2-m_W^2\right)}+
\frac{m_W^2 \log \left(m_W^2\right)}{\left(m_W^2-m_{\tilde{t}}^2\right) \left(m_W^2-m_{\tilde{u}}^2\right)}+
\frac{m_{\tilde{u}}^2 \log \left(m_{\tilde{u}}^2\right)}{\left(m_{\tilde{t}}^2-m_{\tilde{u}}^2\right)
	\left(m_W^2-m_{\tilde{u}}^2\right)}\nn\\
\Omega_{3j}&=&\frac{m_t^4 \log \left(m_t^2\right)}{\left(m_t^2-m_W^2\right) \left(m_t^2-m_{\tilde{u}}^2\right){}^2}+
\frac{m_{\tilde{u}}^4}{\left(m_t^2-m_{\tilde{u}}^2\right){}^2
	\left(m_{\tilde{u}}^2-m_W^2\right)} \left(1+\frac{m_W^2 \log
	\left(m_{\tilde{u}}^2\right)}{m_W^2-m_{\tilde{u}}^2}\right)\label{j3mix}\\
&+&\frac{m_W^4 \log \left(m_W^2\right)}{\left(m_W^2-m_t^2\right) \left(m_W^2-m_{\tilde{u}}^2\right){}^2}+\frac{m_t^2 m_{\tilde{u}}^2 \left(m_W^2 \left(2 \log \left(m_{\tilde{u}}^2\right)+1\right)-m_{\tilde{u}}^2 \left(\log
	\left(m_{\tilde{u}}^2\right)+1\right)\right)}{\left(m_t^2-m_{\tilde{u}}^2\right){}^2
	\left(m_W^2-m_{\tilde{u}}^2\right){}^2}\nn\\
\Omega_{33}&=& \frac{m_{\tilde{t}}^4 \log \left(m_{\tilde{t}}^2\right)}{\left(m_{\tilde{t}}^2-m_t^2\right)
	\left(m_{\tilde{t}}^2-m_{\tilde{u}}^2\right) \left(m_{\tilde{t}}^2-m_W^2\right)}+
\frac{m_W^4 \log \left(m_W^2\right)}{\left(m_W^2-m_t^2\right) \left(m_W^2-m_{\tilde{t}}^2\right)
	\left(m_W^2-m_{\tilde{u}}^2\right)}\nn\\
&+&\frac{m_{\tilde{u}}^4 \log \left(m_{\tilde{u}}^2\right)}{\left(m_{\tilde{u}}^2-m_t^2\right)
	\left(m_{\tilde{u}}^2-m_{\tilde{t}}^2\right) \left(m_{\tilde{u}}^2-m_W^2\right)}+
\frac{m_t^4 \log \left(m_t^2\right)}{\left(m_t^2-m_{\tilde{t}}^2\right) \left(m_t^2-m_W^2\right)
	\left(m_t^2-m_{\tilde{u}}^2\right)}~\cdot\nn
\ea  
We observe that in this simplified limiting case, where all scalar masses are taken equal,
the contributions~(\ref{12gen})  associated with the mixing parameters 
$\xi_{ij}, i,j=1,2$ (where here  $i,j$ are generation indices),  have a very simple 
dependence on the boson mass $m_W$. Contributions involving the third family are
given by~(\ref{j3mix}). 

Notice that the CKM elements multiplying the above contributions are of the same order
$\xi\equiv |\sum_{i,j=1}^2\xi_{ij}|\approx|\sum_{j=1}^2\xi_{3j}|\approx |\xi_{33}|\sim .75\times 10^{-4}$. 
Focusing firstly on the contribution~(\ref{12gen}) of the  two lighter generations,
we observe that the dependence
on the unknown scalar SUSY masses is rather simple, and only ratios of these are
involved. Then, a  rough estimate from contributions coming only from (\ref{12gen})  gives 
\ba
\label{gamma1}
\frac{1}{\tau}=\Gamma \approx -\frac{({\lambda''_{ud\tilde b}})^2 g^4m_{\tilde{b}_{LR}}^{4}
	m_{\tilde{w}}}{32 \pi^{2}(m_{\tilde b_{L}}m_{\tilde{b}_{R}})^{4}}\frac{|\psi(0)|^{4}}{2 m_W^2}\xi~\cdot
\ea
We  assume  equal s-bottom masses $m_{\tilde{b}}=m_{\tilde{b}_{R}}\approx m_{\tilde{b}_{R}}$ and 
define the ratio 
\be c_{b\tilde{b}}\approx m_{\tilde{b}_{LR}}^{2}/m_{\tilde{b}_{R}}^{2}~\cdot \label{bLRmix}
\ee 
Then, we can turn the above expression into
an upper bound for the product $ \lambda_{ud\tilde b}''c_{b\tilde b}$
\be 
 \lambda_{ud\tilde b}''c_{b\tilde b}\le \frac{8\pi m_{\tilde b}^2m_W}{g^2(\tau m_{\tilde W}\xi)^{1/2}}\frac{1}{|\psi (0)|^2}
 ~\cdot 
\ee 
From the lower bound $\tau_{n-\bar n}\gtrsim 10^8$s in free neutron oscillation experiments, for the $\bar n$ annihilation in matter  we can obtain a bound
$\tau_m=\frac{1}{\Gamma_m}> 1.6 \times 10^{31}$y~\cite{Phillips:2014fgb}. 
Assuming the scalar masses to be of the order 
$m_{\tilde{b}}\sim  1$~TeV and taking $m_{\tilde W}=400$ GeV, we obtain 
\be 
{\lambda''_{ud\tilde b}}c_{b\tilde b}\;\lesssim\; 0.5\times 10^{-3}.\label{boundlb}
\ee 

The remaining contributions~(\ref{j3mix}) display a complicated dependence on the SUSY scalar masses, but for masses close to the experimental lower
bounds, they can be of the same order. In such cases, depending on the signs of the CKM mixing parameters $\xi_{jk}$ there might be 
cancellations which result to weaker bounds on the  $\lambda''_{ud\tilde b}$ couplings. 

To examine this general case,  we use the Equation (\ref{GammaW}) to recalculate the bounds on $\lambda''_{ud\tilde  b}$ taken into account the 
latest experimental results for the SUSY mass parameters. We take $A_{eff}=400$ GeV which fixes the ratio~(\ref{bLRmix})
to be $c_{b\tilde b}\sim 10^{-2}$. 

\begin{figure}[!t]
	\centering
	\includegraphics[width=0.6\textwidth,height=0.4\textwidth]{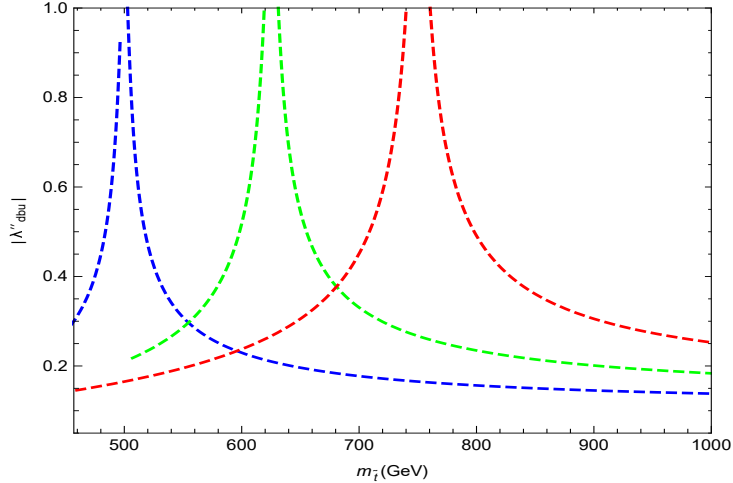}
	\caption{Bounds on $\lambda''_{ud\tilde b}$ for  degenerate up and bottom squark masses 
		$m_{\tilde u}=m_{\tilde c}=m_{\tilde b_L}=m_{\tilde b_R}=450$ GeV. 
		Blue  $m_{\tilde{t}}=500$ GeV, Green $m_{\tilde t}=625$ GeV
		and red $m_{\tilde t}=750$ GeV.}
	\label{A400}
\end{figure}

 In Figure \ref{A400} we fix $m_{\tilde{b}_{L}}= m_{\tilde{b}_{R}}= 450$ GeV.  The three 
 curves correspond to stop masses of 450, 625 and 750 GeV. As we can see, 
 leaving aside accidental cancellations, the value of $\lambda''_{ud\tilde b}$ is constrained to be less that $\sim 0.15-0.3$.

We will now estimate $s_{{\tiny {box}}}$ making use of Eq. (\ref{Eq:exprphi}) by writing:
$$   \frac{1}{m_{\tilde{w}}}s_{{\tiny {box}}}=\frac{1}{3}\frac{1}{32\pi^2}\frac{m_{\tilde{w}}}{m^2_W} (0.75\times 10^{-4})\Rightarrow s_{{\tiny {box}}}\approx 2.0 {\;\rm to \;} 3.0 \times 10^{-6},$$ assuming the range $ m_{\tilde{w}}\approx 400$-$500$ GeV.  We have, of course, removed the factors  3,  (1/2), $g^4$ and the scalar masses from the expression of Eq. (\ref{gamma1}) since they appear explicitly in Eq (\ref{Eq:BoxExpr}).

\end{document}